\documentclass[twocolumn,prl,twoside,superscriptaddress,showpacs,preprintnumbers,floatfix]{revtex4}

\usepackage{amsmath}
\usepackage{amssymb}
\usepackage{graphicx}
\usepackage{dcolumn}
\usepackage{bm}

\def\m#1{\mathrm{#1}}
\def\Eq#1{(\ref{eq:#1})}
\def\d{\mathrm{d}}

\def\epsilon{\varepsilon}
\def\theta{\vartheta}
\def\rho{\varrho}
\def\Int#1#2{\int\!\mathrm{d}^{#1}{#2}\;}

\def\vec#1{\mathbf{#1}}

\begin{document}


\title{Self diffusion of particles in complex fluids: temporary cages and permanent barriers}

\author{Markus Bier}
\email{m.bier@uu.nl}

\author{Ren\'e van Roij}

\affiliation
{
   Institute for Theoretical Physics,
   Utrecht University,
   Leuvenlaan 4,
   3584\,CE Utrecht,
   The Netherlands
}

\author{Marjolein Dijkstra}

\affiliation
{
   Soft Condensed Matter,
   Debye Institute for NanoMaterials Science
   Utrecht University,
   Princetonplein 5,
   3584\,CC Utrecht,
   The Netherlands
}

\author{Paul van der Schoot}

\affiliation
{
   Theoretical and Polymer Physics Group,
   Eindhoven University of Technology,
   Postbus 513,
   5600\,MB Eindhoven,
   The Netherlands
}

\date{6 October, 2008}

\begin{abstract}
   We study the self diffusion of individual particles in dense (non-)uniform complex fluids within dynamic
   density functional theory and explicitly account for their coupling to the temporally fluctuating 
   background particles.
   Applying the formalism to rod-like particles in uniaxial nematic and smectic liquid crystals, we find 
   correlated diffusion in different directions:
   The temporary cage formed by the neighboring particles competes with permanent barriers in periodic 
   inhomogeneous systems such as the lamellar smectic state and delays self diffusion of particles even in
   uniform systems.
   We compare our theory with recent experimental data on the self diffusion of fluorescently labelled 
   filamentous virus particles in aqueous dispersions in the smectic phase and find qualitative agreement.
   This demonstrates the importance of explicitly dealing with the time-dependent self-consistent molecular
   field that every particle experiences.
\end{abstract}


\maketitle


Phenomena such as multi-site hopping in microstructures, void diffusion in colloidal crystals, dynamical
heterogeneities of colloidal glasses and the self-assembly of micelles, supramolecular polymers and viruses
are striking examples of the intriguing dynamics of complex fluids, whose understanding remains relatively
rudimentary despite intense research spanning many decades.
The main reason for this state of affairs presumably is an incomplete understanding of temporal fluctuations
of the fluid structure and how these couple to single-particle diffusive motion.
To circumvent this problem, self diffusion of particles in locally or globally inhomogeneous fluids
is therefore often studied in a fixed background potential mimicking the actual fluid structure but
completely ignoring the inherently fluctuating nature of it \cite{Risken1996}.

It is the aim of this Letter to theoretically investigate the influence of the local fluid structure and the
temporally fluctuating background on the self diffusion in \emph{uniform} and \emph{non-uniform complex fluids}.
As a simple yet quite interesting example we apply our treatment, based on dynamical density
functional theory, to dispersions of elongated colloidal particles in uniaxial nematic and smectic
liquid-crystalline phases.
In the former the orientational degrees of freedom are frozen out and in the latter also one positional degree
of freedom.
This allows us to compare our theory with results of a  very recent experimental study of the
unusual self diffusion in aqueous dispersions of the filamentous bacteriophage \textit{fd} \cite{Lettinga2007}.

As we shall see, the local fluid structure forms a temporary cage around every test particle that
initially hinders its free self diffusion but that decays at later times.
Remarkably, this temporal caging effect of the background particles can produce a coupling between 
motion in different directions in particular if the fluid is symmetry broken.
Both these phenomena influence the self diffusion in structured fluids and cannot be accounted for by
presuming a fixed molecular background field.
In fact, even if a fluctuating molecular field was presumed, its effect can only be predicted beyond the
usual linear analysis of density fluctuations, i.e., they are inherently non-linear.

Focal point of our discussion are the Van Hove correlation functions that probe diffusive processes
\cite{Hansen1986}.
We generalize the formalism introduced for simple fluids in Ref.~\cite{Archer2007}.
The key idea is to define \emph{conditional densities} for which equations of motion can 
be prescribed \cite{Archer2007} and which for non-uniform or complex fluids are non-trivially related 
to the well-known Van Hove correlation functions.
The formalism, although applied here to athermal systems can be generalized straightforwardly to any system
for which a free energy functional can be written down, including, say, thermotropic liquid crystals within
Landau-de Gennes theory \cite{deGennes1995}.


Here, we only outline the main ingredients of the theory.
Consider an equilibrium fluid of $N$ particles with \emph{arbitrary} degrees of freedom
$\vec{x}=(\vec{r},\vec{\omega},\dots)$ and define \emph{self} ($\m{s}$) and \emph{distinct} ($\m{d}$)
\emph{conditional densities} as
\begin{eqnarray}
   C_\m{s}(\vec{x},t|\vec{x'},0)
   & \!\!\!:=\!\!\! &
   \frac{1}{\rho(\vec{x'})}
   \bigg\langle
      \sum_{n=1}^N\delta(\vec{x}-\vec{X}_n(t))\delta(\vec{x'}-\vec{X}_n(0))
   \bigg\rangle,
   \label{eq:CsCddef}\\
   C_\m{d}(\vec{x},t|\vec{x'},0)
   & \!\!\!:=\!\!\! &
   \frac{1}{\rho(\vec{x'})}
   \bigg\langle
      \sum_{\stackrel{\scriptstyle n,n'=1}{n\not=n'}}^N\delta(\vec{x}-\vec{X}_n(t))
      \delta(\vec{x'}-\vec{X}_{n'}(0))
   \bigg\rangle,
   \nonumber
\end{eqnarray}
respectively, where $\langle\cdots\rangle$ is the average over all equilibrium trajectories, $\delta$
the Dirac-$\delta$ in configuration space, $\vec{X}_n(t)$ the configuration of particle $n$ at time $t$, 
and $\rho(\vec{x'})$ the one-particle equilibrium density at configuration $\vec{x'}$.
Note that the distribution of equilibrium trajectories is time-translationally invariant.
The conditional densities $C_\m{s,d}$ and the \emph{Van Hove self} and \emph{distinct correlation
functions} $G_\m{s,d}$ \cite{Hansen1986} are related via
\begin{equation}
   G_\m{s,d}(\Delta\vec{x},t)
   =
   \frac{1}{N}
   \Int{}{x'}C_\m{s,d}(\vec{x'}+\Delta\vec{x},t|\vec{x'},0)\rho(\vec{x'}),
   \label{eq:GbyC}
\end{equation}
where $\vec{x'}+\Delta\vec{x}$ denotes the displacement of configuration $\vec{x'}$ by a suitably defined
offset $\Delta\vec{x}$ in configuration space.
According to Eq.~\Eq{CsCddef}, the conditional densities at zero time $t=0$ read
$C_\m{s}(\vec{x},0|\vec{x'},0) = \delta(\vec{x}-\vec{x'}) =: n_\m{l}(\vec{x},0)$ and
$C_\m{d}(\vec{x},0|\vec{x'},0) = \rho(\vec{x})g(\vec{x},\vec{x'}) =: n_\m{u}(\vec{x},0)$ with $g$ the pair
distribution function \cite{Hansen1986} and $n_\m{l,u}$ the one-particle equilibrium densities of
a fluid of one labeled ($\m{l}$) and $N-1$ unlabeled ($\m{u}$) particles in which the labeled particle is fixed
in configuration $\vec{x'}$.
Upon releasing the fixed labeled test particle at time $t=0$, the neighboring host fluid relaxes towards a
new equilibrium state and it is assumed that the time-dependent two-particle correlators
$C_\m{s,d}(\vec{x},t|\vec{x'},0)$ equal the one-particle densities $n_\m{l,u}(\vec{x},t)$ for \emph{all} times
$t \geq 0$.
This identification is reminiscent of Onsager's regression hypothesis and implies the fluctuation-dissipation
theorem to hold, which is strictly proven only within linear response theory
\cite{MariniBettoloMarconi2008}.
A free energy density functional $F[n_\m{l},n_\m{u}]$ \cite{Evans1979} describing
the fluid automatically produces the one-particle equilibrium density distribution $\rho$ and
the pair distribution function $g$ as well as expressions for the \emph{local chemical potentials}
\begin{equation}
   \mu_\m{l,u}(\vec{x},t,[n_\m{l},n_\m{u}])
   :=
   \frac{\delta F}{\delta n_\m{l,u}(\vec{x})}\bigg|_{n_\m{l}(\cdot,t),n_\m{u}(\cdot,t)}
   \label{eq:locchempot}
\end{equation}
that within a generalized Fickian approximation give rise to the equations of motion \cite{Dieterich1990}
\begin{eqnarray}
   & &
   \frac{\partial}{\partial t}n_\m{l,u}(\vec{x},t) =
   \label{eq:eom}\\
   & &
   \nabla_\vec{x}\cdot(n_\m{l,u}(\vec{x},t)\Gamma(\vec{x},t,[n_\m{l},n_\m{u}])
   \cdot\nabla_\vec{x}\mu_\m{l,u}(\vec{x},t,[n_\m{l},n_\m{u}]))
   \nonumber
\end{eqnarray}
with $\Gamma$ the mobility matrix.
Note that the Fickian approximation leading to Eq.~\Eq{eom} implies the hypothesis of local equilibrium and the
neglect of hydrodynamic interactions, i.e., the free draining limit to effectively hold.
In dense systems the latter presumption seems reasonable due to screening of hydrodynamics.
The mobilities are then to be interpreted as renormalized ones.
Integrating this closed set of equations for $n_\m{l,u}$ subject to the specified initial conditions and
interpreting $n_\m{l,u}$ as conditional densities $C_\m{s,d}$, one obtains the Van Hove correlation
functions $G_\m{s,d}$ from Eq.~\Eq{GbyC}.
For a \emph{uniform, simple} fluid, where the particles possess only translational degrees of freedom,
$\vec{x}=\vec{r}$, and the one-particle density $\rho$ is spatially constant, the conditional densities
$C_\m{s,d}(\vec{r},t|\vec{r'},0)$ depend only on $\vec{r}-\vec{r'}$, so that
$G_\m{s,d}(\Delta\vec{r},t) = C_\m{s,d}(\Delta\vec{r},t|\vec{0},0) = n_\m{l,u}(\Delta\vec{r},t)$ and
Eqs.~\Eq{locchempot} and \Eq{eom} can be interpreted as equations of motion for the Van Hove correlation functions
$G_\m{s,d}$ and avoids having to explicitly deal with kinetic equations for conditional 
densities that in actual fact are two-point correlators.
Note that the quantities called ``van Hove functions'' in Ref.~\cite{Archer2007} are actually conditional 
densities, which, in general, differ from the well-known Van Hove correlation functions.
The general conditional density formalism described here is applicable to any \emph{non-uniform} or
\emph{complex} fluid and offers a route to analyze the relaxational dynamics of a wide range of
interesting complex fluids.


Motivated by the very recent measurements of the Van Hove self correlation function of aqueous solutions of
the bacteriophage \textit{fd} \cite{Lettinga2007}, a filamentous virus particle of about $900\,\m{nm}$ length
and $7\,\m{nm}$ width, we consider a free energy functional $F[n_\m{l},n_\m{u}]$ describing a lyotropic liquid
crystal of (stiff) hard rods of length $L$ and diameter $D$ with $L \gg D$.
As the root-mean-squared angle between the axis of a rod and the director scales as $\mathcal{O}(D/L)$ in this limit
\cite{Odijk1986}, the orientational degrees of freedom can be ignored, so the rods are assumed to be
oriented parallel to the director, which itself is assumed to be fixed.
This means that the model particles possess only translational and no orientational or internal conformational
degrees of freedom such as arising from a bending flexibility.
For the free energy density $F[n_\m{l},n_\m{u}]$ we for reasons of simplicity invoke the second virial
approximation \cite{Mulder1987}.
Although not accurate it is known to capture the main features of the structure of the smectic phase near the
nematic transition point, which suffices for our purposes.
Our model of perfectly parallel hard rods is simple but not overly simple, as quantitative precision is
not required and the neglect of features such as particle flexibility, higher virial terms, or (screened)
hydrodynamic interaction can in principle be accounted for by renormalization of the model parameters.
The bare \emph{translational mobilities} $\Gamma_\|$ and $\Gamma_\perp$ parallel and perpendicular to the 
director give rise to the \emph{parallel translational diffusion time} $\tau:=\beta L^2/\Gamma_\|$ and the 
\emph{diffusion rate ratio} $\gamma := \frac{\Gamma_\perp/D^2}{\Gamma_\|/L^2}$.

\begin{figure}[t]
   \includegraphics[width=8cm]{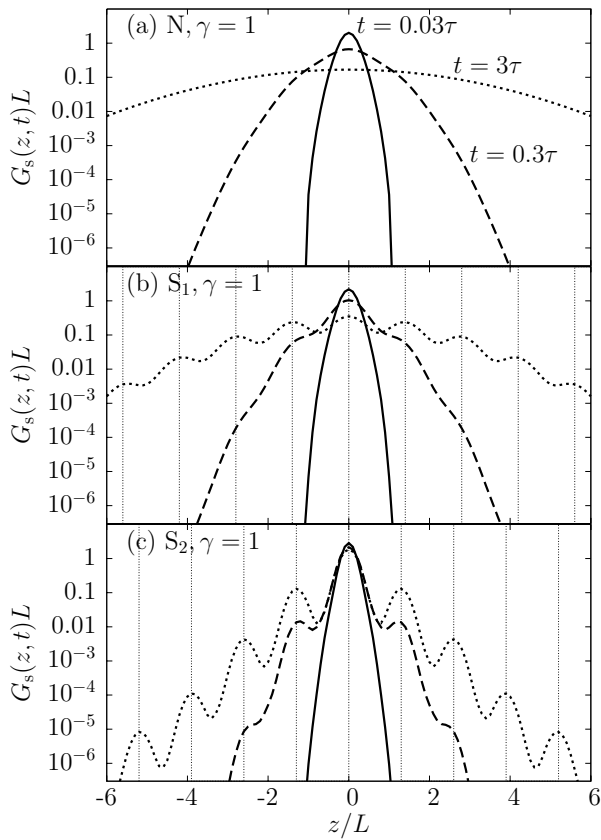}
   \caption{\label{fig:1}Radially integrated Van Hove self correlation function $G_\m{s}(z,t)$ as a function of the
           axial displacement $z$ in units of the rod length $L$ and time $t$ in units of the diffusion time $\tau$
           for diffusion rate ratio $\gamma=1$ for (a) the nematic state $\m{N}$, (b) the weakly smectic state
           $\m{S_1}$, and (c) the strongly smectic state $\m{S_2}$.
           The thin vertical lines in (b) and (c) indicate the centers of the smectic layers.}
\end{figure}

\begin{figure}[t]
   \includegraphics[width=8cm]{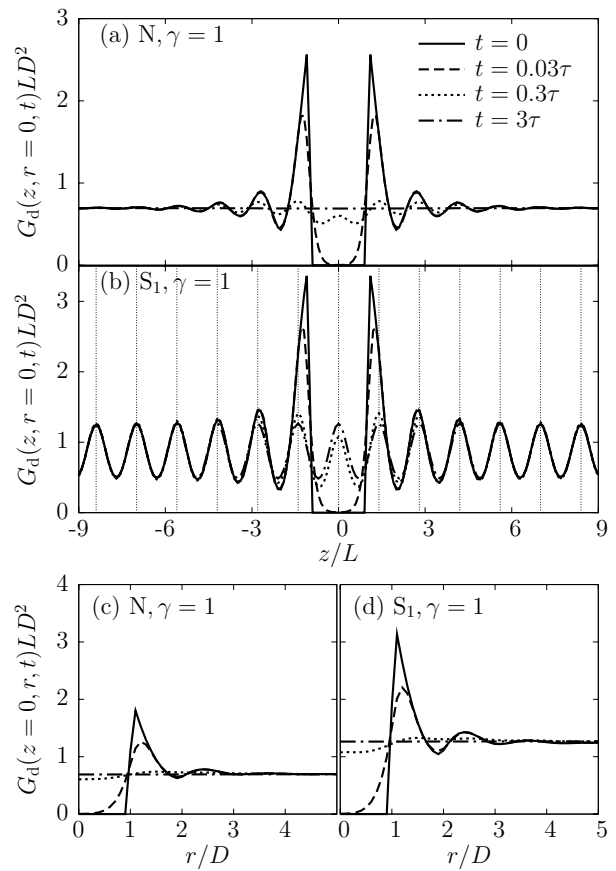}
   \caption{\label{fig:2}Cuts of the Van Hove distinct correlation function $G_\m{d}(z,r,t)$ as a function of the
           axial displacement $z$ in units of the rod length $L$, the radial displacement $r$ in units of the
           rod diameter $D$, and time $t$ in units of the diffusion time $\tau$ for diffusion rate ratio
           $\gamma=1$ for (a,c) the nematic state $\m{N}$ and (b,d) the weakly smectic state $\m{S_1}$.
           The corresponding curves for the strongly smectic state $\m{S_2}$ are similar to those in (b,d).
           The thin vertical lines in (b) indicate the centers of the smectic layers.}
\end{figure}

\begin{figure}[t]
   \includegraphics[width=8cm]{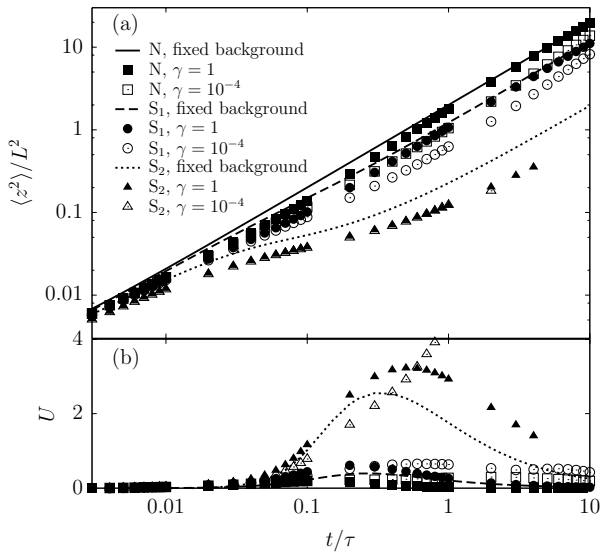}
   \caption{\label{fig:3} (a) Parallel mean-squared displacement $\langle z^2 \rangle$ in units of the squared
           rod length $L^2$ and (b) Binder cumulant $U$ (see main text) of the labeled particle
           as a function of time $t$ in units of the diffusion time $\tau$ for the diffusion rate ratios
           $\gamma=1$ and $\gamma=10^{-4}$ for the nematic state $\m{N}$, the weakly smectic
           state $\m{S_1}$ and the strongly smectic state $\m{S_2}$.
           The points represent the results obtained within the present fluid model, whereas the lines describe
           the one-dimensional self diffusion in a temporally fixed background potential corresponding to the
           equilibrium one-particle density.}
\end{figure}

We calculated the Van Hove correlation functions $G_\m{s,d}(z,r,t)$ as a function
of the parallel displacement $z$, the perpendicular displacement $r$, and the time $t$.
Results for the largely arbitrary but definitely representative choice of parameters $\gamma=1$
(representing fast radial diffusion) and $\gamma=10^{-4}$ (slow radial diffusion) for the three cases of a
nematic state $\m{N}$ at a chemical potential $\beta\Delta\mu=-0.388$ relative to that at the nematic-smectic
transition, a weakly  smectic state $\m{S_1}$ with $\beta\Delta\mu=0.612$, i.e., at a density just above the
smectic transition, and a strongly smectic state $\m{S_2}$ with $\beta\Delta\mu=2.612$ are shown in
Figs.~\ref{fig:1}, \ref{fig:2}, and \ref{fig:3}.
In passing we note that parallel diffusion becomes independent of the parameter $\gamma$ if
$\gamma \geq 1$, because perpendicular relaxation is then much faster than parallel relaxation according to 
our numerical results.
The interlayer distances of the smectic phases within the present model are $1.4L$ for $\m{S_1}$ and $1.3L$ for 
$S_2$; these values would be somewhat smaller had we included particle flexibility \cite{vanderSchoot1996} and 
higher virial terms \cite{Mulder1987}.
For these two smectic states, the calculated lamellar density undulations correspond to a self-consistent
molecular field barrier height of $2.1k_BT$ and $4.7k_BT$, respectively.


The radially integrated Van Hove self correlation functions $G_\m{s}(z,t):=2\pi\int_0^\infty\!\d r\,rG_\m{s}(z,r,t)$
presented in Fig.~\ref{fig:1} exhibit a spatial broadening due to self diffusion, which is slower the larger the
background potential barrier height is.
Whereas  $G_\m{s}(z,t)$ is a concave function of $z$ for the nematic state $\m{N}$ for all times $t$, shoulder
peaks develop in the smectic states $\m{S_1}$ and $\m{S_2}$ located at the centers of the smectic layers in
accord with the experimental findings of Ref.~\cite{Lettinga2007}.
These peaks are the manifestation of the existence of the average self-consistent field due to the equilibrium
one-particle density.
The curves of the \emph{nematic} state and the envelopes to the curves of the smectic states are \emph{not}
Gaussians (see Fig.~\ref{fig:3}b) because of the influence of the background fluid that cages the
test particle.
Upon approaching the nematic-smectic phase transition from the high-density, smectic side we found an increase
of the time that it takes for the first shoulder peak to appear (see also the curves $t=0.3\tau$ in 
Figs.~\ref{fig:1}b and c).
This we attribute to influence of the critical slowing down of the collective dynamics of the host fluid when
nearing the smectic spinodal. 
This may well explain the absence of shoulder peaks in Fig.~2b of Ref.~\cite{Lettinga2007} presuming that this 
time exceeds the time of measurement.


Figure~\ref{fig:2} shows the time-evolution of the structure of the fluid background for the nematic
and the smectic phase in terms of the Van Hove distinct correlation function $G_\m{d}(z,r,t)$.
The initial Van Hove distinct correlation function $G_\m{d}(z,r,t=0)$ coincides with the pair distribution
function and the nearest-neighbor (solvation) shell in the vicinity of the origin corresponds to a cage around
the labeled test particle.
This cage initially slows down the self diffusion of the labeled test particle, but it dissolves with time.
Eventually, the labeled particle diffuses effectively only in the permanent background of the unlabeled particles.
The presence of the temporary cage that adds to the permanent equilibrium background barriers between the
particle layers is a direct consequence of the local fluid structure, which cannot be accounted for by fixed
background models as in Ref.~\cite{Risken1996}.
For $\gamma=1$, the relaxation of $G_\m{d}(z,r,t)$, i.e., the decay of the temporary cage, takes place on a 
time scale of the diffusion time $\tau$.
If the perpendicular diffusion is strongly suppressed, e.g., for $\gamma=10^{-4}$, this relaxation
takes place on a larger time scale, indicating that the dissolution of the temporary cage is linked to the
perpendicular diffusion.
Hence, the time-dependent fluid background mediates a coupling between parallel and perpendicular self diffusion.


The influence of the fluid structure on the self diffusion of individual test particles can be inferred from
Fig.~\ref{fig:3} that shows our results for their mean-squared displacement $\langle z^2 \rangle$ along the
director of the fluids and for the Binder cumulant 
$\displaystyle U:=\langle z^4 \rangle / (3\langle z^2 \rangle^2) - 1$, which  quantifies deviations from 
Gaussian behavior \cite{Risken1996}, where $\langle z^k \rangle :=\int\!\d z\,z^k G_\m{s}(z,t)$. 
The significant difference between results obtained within fixed background models \cite{Risken1996}
and our fluid background model highlight the importance of the temporal nature of fluid structure in general
and of the sideways diffusion in the intermediate-time regime of uniaxial fluids in particular.
For each of the three states investigated, one spatially homogeneous and two spatially inhomogeneous, caging
causes diffusion initially to be slowed down giving rise to a non-vanishing Binder cumulant.
The crossover from early to late-stage diffusive motion of test particles in a congested fluid
causes this deviation from Gaussian behavior and can be linked to potential barriers that can be temporary or 
fixed.
In either case, they are due to correlations between particles in the fluid that cannot reasonably be
described by a fixed potential.
Our calculations show that in the crossover region sub-diffusive behavior may be inferred over a limited time
range, as done in Ref.~\cite{Lettinga2007}.
The slowing down of \textit{fd} virus particle diffusion in the smectic phase was rationalized in
Ref.~\cite{Lettinga2007} by considerations involving a fixed periodic molecular field only.
In our view this does not do justice to the complexity of the problem that involves a coupling of
between-layer and in-layer diffusive processes. 
This can be inferred from a comparison of the cases $\gamma=1$ and $\gamma=10^{-4}$ for the states $\m{N}$
and $\m{S_1}$ where a reduction of the \emph{perpendicular} diffusivity gives rise to a slower 
\emph{parallel} self diffusion.
On the other hand, the mean-squared displacement for strongly smectic states for which the permanent potential
barriers are quite large is virtually independent of $\gamma$, as the initial delay of the diffusion is
due to the high, $\gamma$-independent permanent barriers and not due to the temporal cage.


In summary, we found a remarkable influence of the fluctuating local structure on the self diffusion of
particles in spatially (non-)uniform complex fluids.
Diffusion delays initially even in uniform fluids due to a temporary cage formed by neighboring particles.
This coupling of the motion of a test particle to its surrounding particles causes motion in different 
directions to become coupled, in particular in anisotropic fluids.
Fixed molecular background models cannot describe this and have to be replaced by more sophisticated
approaches such as the dynamic density functional formalism presented here.


\begin{acknowledgments}
We thank M.\ P.\ Lettinga, E.\ Grelet, and B.\ Vorselaars for useful comments as well as
M.\ Schmidt, A.\ J.\ Archer, and P.\ Hopkins for a clarifying communication. 
This work is part of the research program of the ``Stichting voor
Fundamenteel Onderzoek der Materie (FOM)'', which is financially supported by
the ``Nederlandse Organisatie voor Wetenschappelijk Onderzoek (NWO)''.
\end{acknowledgments}



\end{document}